\documentclass{jfm}
\usepackage{graphicx}
\usepackage{amsmath}
\usepackage{derivative}

\newcommand{\laplacian}[1]{\nabla^2 #1}
\newcommand{\cz}{c_{\zeta}}
\newcommand{\cs}{c_{\psi}}
\newcommand{\ct}{c_{\theta}}
\newcommand{\cu}{c_u}
\newcommand{\css}{c_{\psi \psi}}
\newcommand{\csz}{c_{\psi \zeta}}
\newcommand{\czs}{c_{\zeta \psi}}
\newcommand{\czz}{c_{\zeta \zeta}}
\newcommand{\ctz}{c_{\theta \zeta}}
\newcommand{\czt}{c_{\zeta \theta}}
\newcommand{\ctt}{c_{\theta \theta}}
\newcommand{\cst}{c_{\psi \theta}}
\newcommand{\cts}{c_{\theta \psi}}
\newcommand{\Rayleigh}{\mbox{\textit{Ra}}}  

\shorttitle{CE2 Busse Annulus}
\shortauthor{J. S. Oishi, K. J. Burns, J. B. Marston, S. M. Tobias}

\title{Direct Statistical Simulation of the Busse Annulus}

\author{Jeffrey S. Oishi\aff{1}
  \corresp{\email{joishi@bates.edu}},
  Keaton J. Burns\aff{2,3},
  J. B. Marston\aff{4},
 \and Steven M. Tobias\aff{5}}

\affiliation{\aff{1}Department of Physics \& Astronomy, Bates College,
Lewiston, ME 04240, USA
\aff{2} Department of Mathematics, Massachusetts Institute of Technology, Cambridge, MA 02138 USA
\aff{3} Center for Computational Astrophysics, Flatiron Institute, New York, NY 10010, USA
\aff{4} Department of Physics and Brown Theoretical Physics Center, Brown University, Providence, RI 02912, USA
\aff{5} Department of Applied Mathematics, University of Leeds, Leeds LS2 9JT, UK
}

\begin{document}

\maketitle

\begin{abstract}
We consider direct statistical simulation (DSS) of a paradigm system of convection interacting with mean flows. In the Busse Annulus model zonal jets are generated through the interaction of convectively driven turbulence and rotation; non-trivial dynamics including the emergence of multiple jets and bursting `predator-prey' type dynamics can be found. We formulate the DSS by expanding around the mean flow in terms of equal-time cumulants and arrive at a closed set of equations of motion for the cumulants. Here, we present results using an expansion terminated at the second cumulant (CE2); it is fundamentally a quasi-linear theory.
We focus on particular cases including bursting and bistable multiple jets and demonstrate that CE2 can reproduce the results of direct numerical simulation if particular attention is given to symmetry considerations. 
\end{abstract}


\section{Introduction}
\label{sec:intro}

Turbulence interacts with, and leads to the generation of, mean flows in a wide variety of natural systems.
Understanding the nature of these interactions, particularly for systems far from equilibrium, remains a key problem for the description of large-scale flows on planets and stars.
The fundamental problem in studying such systems via direct numerical simulation (DNS) is the expense of the calculations, owing to the vast range of spatial and temporal scales that need to be described. 
One way to make progress is to study the statistics of the flow rather than detailed flow variables themselves \citep{marston_qi_tobias_2019}. However, one should bear in mind that these systems are often inhomogeneous and strongly anisotropic, and such a statistical description should respect this, despite the simplifications that are afforded with assumptions of homogeneity and isotropy.

Rotation is a hallmark of geophysical and astrophysical fluid dynamics.
Through the breaking of reflection symmetry, rotating systems generate mean flows such as zonal jets; examples of which include the bands on Jupiter, the jet stream on Earth and the differential rotation in stars \citep{galperin_read_2019}.
In these systems, turbulence that is driven at moderate scales interacts with rotation to produce non-trivial correlations that lead to Reynolds stresses that drive differential rotation and jets. Here, the turbulence is often described as having an anti-frictional character as it moves the system \textit{away} from solid body rotation; this provides a stern test for theories that seek to describe the statistical properties of the turbulence. We also note that it is common astrophysically for the source of the turbulence to be thermal convection, and that the convection itself is sensitive to the presence or absence of large-scale flows; this leads to non-trivial feedbacks between the mean flows and the turbulence.

Here, we explore the simplest system that describes the interaction of convection with mean flows --- the Busse Annulus. The Busse Annulus models rotating convection in an annulus with slanted ends leading to a topographic $\beta$ effect; this effect leads to vortex stretching and the generation of systematic large-scale flows 
\citep[see e.g.][]{1976Icar...29..255B,bh1993,rj2006}. We use this system to test the effectiveness of the statistical description termed CE2. CE2 is a quasilinear approximation of direct statistical simulation (DSS), and is a cumulant expansion truncated at second order. CE2 has been shown to be effective for simple systems where tightly coupled correlations control the dynamics and driving is either stochastic or arises through the instability of a shear flow. However the case of thermal convection, where buoyancy provides the driving in the vorticity equation and there is nonlinearity in both the vorticity and temperature equations, has been much less studied. 

Thus, the Busse Annulus system presents an important challenge for CE2, not least because it is known to host multiple solutions at modest Rayleigh numbers \citep{bh1993}.
How does a quasilinear statistical theory fare when faced with multiple potential solution basins?
In particular, we investigate the sensitivity of CE2 to initial conditions for a case with multiple solution branches.

\section{Model Equations, Formulation, Parity and Numerical Methods}
\label{sec:model-eqations}

The Busse Annulus we consider  is a locally Cartesian model of rotating (at angular velocity ${\boldsymbol \Omega = \Omega {\bf e}_z}$), incompressible Boussinesq fluid with viscosity $\nu$ and thermal diffusivity $\kappa$. Here, gravity is uniform and in the $y$-direction. 
The system is non-dimensionalised, with a fiducial length being the width of the annulus in the $y$-direction ($d$), a typical timescale being the viscous timescale $d^2/\nu$ and a typical velocity scale $\nu/d$; a typical temperature is chosen to be $\Delta T$, the temperature difference between the inner and outer walls. 

Following \citet{bh1993, rj2006}, 
the temperature  $T = T_{BS} +{\hat \theta(x,y)}$ is decomposed into a basic state profile $T_{BS}$ satisfying $\nabla^2 T_{BS} = 0$ and a perturbation ${\hat \theta}$. 
After some manipulation, the equation for the vertical ($z$) component of the vorticity ($\zeta$) is given by
\begin{equation}
  \label{eq:zeta_eom}
  \pdv{\zeta}{t} + J(\psi, \zeta) - \beta \pdv{\psi}{x} = -\frac{\Rayleigh}{\Pran} \pdv{\theta}{x} -C |\beta|^{1/2} \zeta + \laplacian{\zeta},
\end{equation}
where $\psi$ is the streamfunction, $(u,v) = (-\pdv*{\psi}{y}, \pdv*{\psi}{x})$, and $\zeta$ is related to $\psi$ by
\begin{equation}
  \label{eq:zeta_def}
  \zeta = \laplacian{\psi}.
\end{equation}
Here $J(A, B) = \pdv*{A}{x}\pdv*{B}{y} - \pdv*{A}{y}\pdv*{B}{x}$ is the Jacobian. 
The equation for the temperature perturbation  $\theta$ is given by
\begin{equation}
  \label{eq:theta}
  \pdv{\theta}{t} + J(\psi, \theta) = -\pdv{\psi}{x} + \frac{1}{\Pran} \nabla^2 \theta.
\end{equation}
The system is governed by four dimensionless parameters, $\beta$, $C$, $\Pran$, and $\Rayleigh$ \citep[see e.g.][for the definition of these parameters]{tom_2018}. Here $\beta$ measures the degree of vortex stretching engendered by the sloping endwalls and $C$ measures the degree of friction, whilst the more familiar Rayleigh number and thermal Prandtl number have their usual physical interpretations.

A zonal CE2 formulation (sometimes termed zCE2) is implemented using a zonal average, 
\begin{equation}
\langle f(x,y) \rangle = \frac{1}{L_x} \int_0^{L_x} f(x,y) dx
\end{equation}
to split all dynamical variables into mean and fluctuating components,
\begin{equation}
    f(x,y,t) = \langle f \rangle + f'.
\end{equation}
We then derive evolution equations for the first cumulants $\cz(y) = \langle \zeta \rangle $ and $\ct(y) = \langle \theta \rangle$ and the second cumulants 
\begin{equation}
    \ctt(\xi,y_1,y_2) = \langle \theta^\prime(x_1,y_1) \theta^\prime(x_2,y_2) \rangle
\end{equation}
with $\xi = x_1 - x_2$.
Similar definitions are used for the second cumulants $\czz$, $\ctz$, and $\czt$. 
Cumulants involving $\psi$ and $\zeta$ are related by differential operators \citep{2013PhRvL.110j4502T}.
Because the equations require both $\psi$ and $\zeta$ cumulants, we solve for $\cs$, $\ct$, $\css$, $\cts$, $\cst$, and $\ctt$, though we write equations in terms of $\zeta$ cumulants.

The  dynamical equations for the first cumulants are given by
\begin{equation}
  \label{eq:cz}
  \pdv{\cz}{t} = - \left(\pdv{}{y_1} + \pdv{}{y_2}\right) \pdv{\csz}{\xi}\Big|_{\xi \to 0}^{y_1 \to y_2} - C |\beta|^{1/2} \cz + \pdv[2]{\cz}{y_1}
\end{equation}
and
\begin{equation}
  \label{eq:ct}
  \pdv{\ct}{t} = - \left(\pdv{}{y_1} + \pdv{}{y_2}\right) \pdv{\cst}{\xi} \Big|_{\xi \to 0}^{y_1 \to y_2} + \frac{1}{\Pran} \pdv[2]{\ct}{y_1},
\end{equation}
whilst those for the second cumulants are given by
\begin{equation}
  \label{eq:czz}
  \begin{split}
    \pdv{\czz}{t} &= \pdv{\cs}{y_1} \pdv{\czz}{\xi} - \left(\pdv{\cz}{y_1} - \beta\right) \pdv{\csz}{\xi} - \pdv{\cs}{y_2} \pdv{\czz}{\xi}  + \left(\pdv{\cz}{y_2} - \beta\right) \pdv{\czs}{\xi}\\
    &+ \frac{\Rayleigh}{\Pran} \left(\pdv{\czt}{\xi} -  \pdv{\ctz}{\xi}\right) - 2C |\beta|^{1/2} \czz + (\nabla_1^2 + \nabla_2^2) \czz,    
  \end{split}
\end{equation}

\begin{equation}
  \label{eq:ctt}
\begin{split}
  \pdv{\ctt}{t} &= \pdv{\cts}{\xi} - \pdv{\cst}{\xi} + \left(\pdv{\cs}{y_1} - \pdv{\cs}{y_2}\right)\pdv{\ctt}{\xi} + \pdv{\ct}{y_2}\pdv{\cts}{\xi} - \pdv{\ct}{y_1}\pdv{\cst}{\xi}\\
&  + \frac{1}{\Pran}(\nabla_1^2 + \nabla_2^2) \ctt,
\end{split}
\end{equation}
and
\begin{equation}
  \label{eq:ctz}
  \begin{split}
    \pdv{\ctz}{t} &= \left(\pdv{\cs}{y_1} - \pdv{\cs}{y_2}\right) \pdv{\ctz}{\xi} - \left(1 + \pdv{\ct}{y_1}\right) \pdv{\csz}{\xi} + \left(\pdv{\cz}{y_2} - \beta\right) \pdv{\cts}{\xi}\\
    &  - C |\beta|^{1/2} \ctz + \frac{1}{\Pran} \nabla_1^2 \ctz + \nabla_2^2 \ctz.
  \end{split}
\end{equation}
The other second cumulant terms (e.g. $\czt$) do not need their own evolution equations, as they can be calculated from symmetry considerations.

We solve both the dynamical PDE (DNS) system and the cumulant system subject to impenetrable, stress-free, zero temperature perturbation boundary conditions in the $y$ dimension; both systems are periodic in $x$. 
This implies that $\theta$, $\zeta$, and $\psi$ all have odd parity in $y$ (and are periodic in $x$). As
the action of the zonal average preserves the parity, we discretise the first cumulants using a sine series in $y$. Likewise,
the second cumulants may be discretised using sine series in $y_1$ and $y_2$ and a Fourier basis in $x$.

We use \emph{Dedalus} \citep{2020PhRvR...2b3068B} to solve both the direct equations (\ref{eq:zeta_eom} - \ref{eq:theta}) and the CE2 model equations, (\ref{eq:cz})-~(\ref{eq:ctz}).
DNS use the SBDF2 time-stepper with linear terms implicit, while
the CE2 equations use RK222 scheme with all terms evaluated explicitly.
We have found fully explicit timestepping to be more stable for CE2.
The second cumulants must be positive definite in order to ensure the corresponding probability density function is realisable.
At higher $\Rayleigh$, we have found it necessary to remove negative eigenvalues from the second cumulant every 10 timesteps.
While this is required for CE2.5 \citep{marston_qi_tobias_2019}, it is also necessary in these CE2 simulations.

\section{Results}
\label{sec:results}
Given that the Busse Annulus system is known to exhibit strong hysteresis, it is key to examine the role of initial conditions when solving the cumulant system.
Minimally, we begin each CE2 simulation using one of two extreme initial conditions: \emph{maximum ignorance}, in which we assume a completely uncorrelated $\ctt$ and all other cumulants (both first and second) zero, and \emph{maximum knowledge}, in which we initialise all fields using cumulants calculated from the statistics of a DNS solution. This allows the evaluation of CE2's ability to both \emph{find} solutions and \emph{continue} solutions already known. In addition, we begin some solutions with biased initial conditions that are designed to ensure solutions are not trapped in symmetry subspaces.
In order to facilitate comparison with previous work, we adapt the same run naming scheme as \citet{tom_2018}.

\begin{table}
  \begin{center}
\def~{\hphantom{0}}
  \begin{tabular}{lcccccl}
    Run & $\beta$ & $\Rayleigh$ & $C$ & DNS Resolution & CE2 Resolution & Comment \\[3pt]
    A  & $2.8\times 10^3$ & $7.6 \times 10^4$   & $0$ & $256 \times 64$ & $32^3$ & Two-jet \\
    B  & $7.07\times 10^5$ & $10^8$ & $0.316$ & $512 \times 256$ & $128 \times 64$ & Seven-jet  \\
    C  & $5\times 10^5$ & $8\times 10^7$ & $0$ & $512 \times 256$ & $256 \times 64$ & three-jet $\to$ two-jet bursting \\
    R  & $3.16\times 10^5$ & $4\times 10^7$ & $0.316$ & $256 \times 64$ & $128\times 64^2$ & Five-jet\\
    Rb  & $3.16\times 10^5$ & $4\times 10^7$ & $0.316$ & - & $128\times 64^2$ & CE2 3-jet initial bias\\
  \end{tabular}
  \caption{Runs}
  \label{tab:runs}
  \end{center}
\end{table}

\subsection{Run A: 2 Jet/3 Jet solutions}
\label{sec:hov}

\begin{figure}
  \centering
  \includegraphics[width=\textwidth]{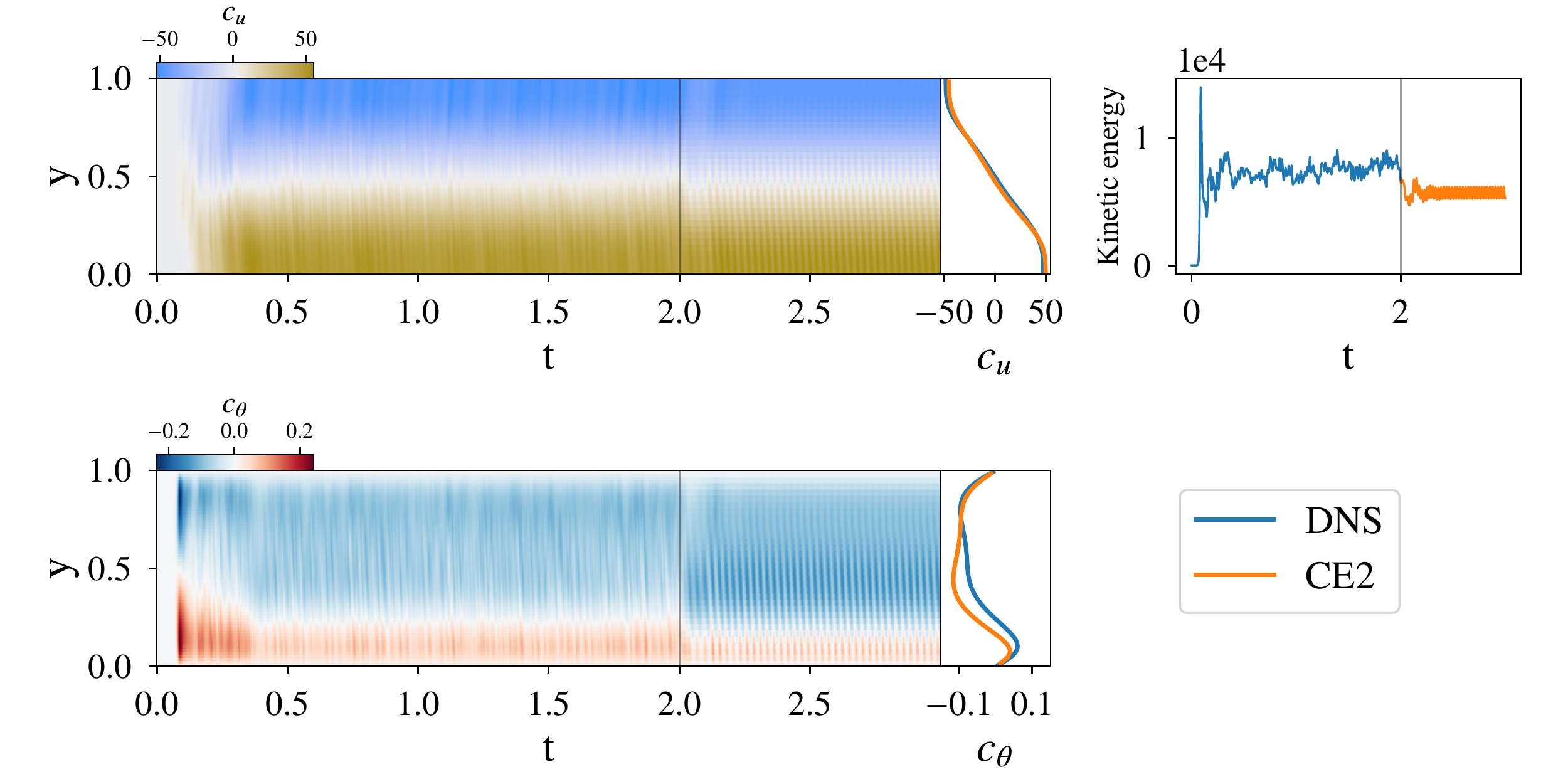}
  \caption{Left: Hovm\"oller diagrams of first cumulants $\cu$ (top), $c_\theta$ (bottom) from run A started in DNS and continued with CE2 at $t=2$. Attached to the right $y$ axis of the Hovm\"oller diagrams: first cumulants of $\cu$ (top) and $\ct$ (bottom) as a function of $y$ for DNS averaged over $1 \le t \le 2$ and CE2 averaged from $2 \le t \le 3$. Right: total kinetic energy.}
  \label{fig:run_A}
\end{figure}

The first case we consider (case A) exhibits relatively simple dynamics in DNS. Started from random initial conditions, the driven convection interacts with the $\beta$-effect to drive large-scale zonal flows (jets). These takes the form of one prograde and one retrograde jet as shown in the solutions for the mean zonal velocity $\langle u \rangle(y, t)$ and mean temperature $\langle \theta\rangle(y,t) $ in the Hovm\"oller plots of Figure~\ref{fig:run_A}. 
Started from maximum ignorance initial conditions, however, CE2 evolves to a 3-jet solution, which is symmetric about the midplane ($y=0.5$). These results are similar to the strictly quasilinear run in \citet{tom_2018}, which also produces a 3-jet solution; an unsurprising result given that 
 CE2 is fundamentally a quasi-linear theory. Interestingly, generalized quasilinear (GQL) solutions do however yield the correct number of jets \citet{tom_2018}. At these parameters it has been demonstrated in DNS that there is hysteresis between 2-jet and 3-jet solutions, but that 3-jet solutions are unstable to perturbations that break shift-reflect symmetry \citep{bh1993}. To test this, we ran a DNS with initial conditions that respect this symmetry and, owing to our highly accurate spectral methods, the solution remained in the symmetry class and the 3-jet solution remained stable in DNS.

Better comparison with DNS for these parameter values is achieved when the CE2 solution is biased 
 by initialising the first cumulant of the x-velocity, $\cu$ to
\begin{equation}
  \label{eq:bias}
  \cu(t=0, y_0) = A_0 \left( \lambda \cos(2\pi/L_y y_0) + (1-\lambda) \cos (\pi/L_y y_0)\right).
\end{equation}
The first term is odd about the center of the domain, while the second is even.
We find that CE2 \emph{is} capable of finding and sustaining a 2-jet solution, with amplitude comparable with the DNS results, if the symmetry of the initial condition is sufficiently biased.
This strongly suggests that the transition from 3 to 2 jets in DNS is the product of a subcritical, non-linear transition mediated by eddy-eddy $\to$ eddy interactions excluded both from CE2 and our previous quasilinear results.
Regardless, this solution remains a fixed point of the CE2 system, suggesting that the \emph{selection} of a given multi-jet solution is distinct from its \emph{maintenence}.
This hypothesis is supported by our ``maximal knowledge" calculation where the CE2 solution is initialised with the cumulants calculated from the saturated state of the DNS run (see Figure~\ref{fig:run_A}). CE2 is certainly capable of maintaining the form of these fully nonlinear solutions (even maintaining the slight asymmetry in the temperature distribution) even though the eddy-eddy non-linearity (EENL) term is missing from the system.
\begin{figure}
    \centering
    \includegraphics[width=\textwidth]{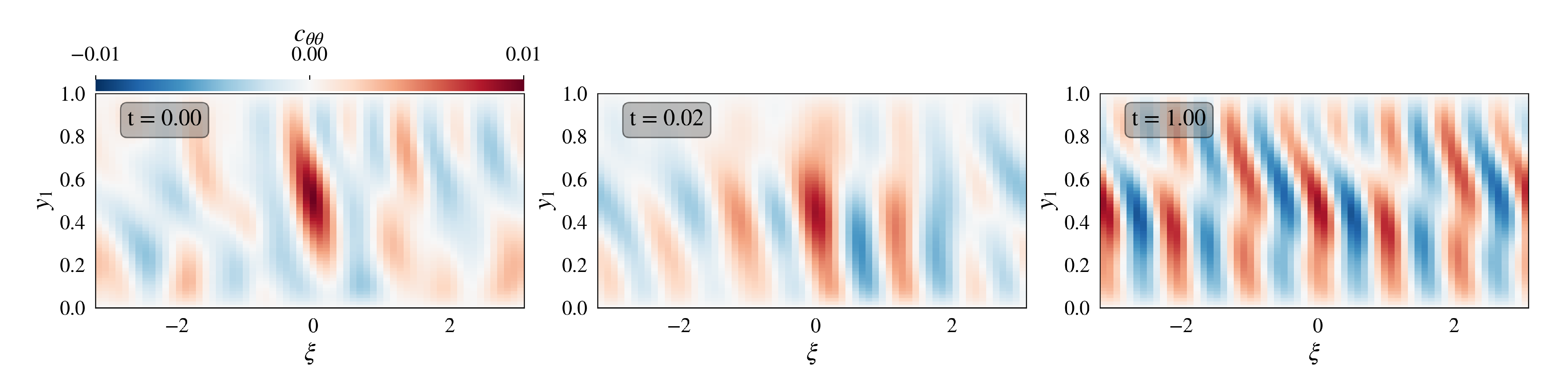}
    \caption{$\ctt(\xi, y_1, y_2 = 0.5)$ at three different times for run A with ``maximum knowledge'' initial conditions selected from the output of DNS. The initial, tightly peaked $\ctt$ from DNS (left most panel) quickly delocalises and reverts to the overemphasis of long-range correlations characteristic of QL models including CE2.}
    \label{fig:run_A_decoherence}
\end{figure}

The degree of success of the quasilinear CE2 models can be investigated further by comparing the second cumulants; it is possible for first cumulants to agree well whilst second cumulants diverge. Figure~\ref{fig:run_A_decoherence} shows the evolution of the second cumulant $\ctt(\xi, y_1, y_2 = 0.5)$ under CE2 for the solution started from the statistics of the saturated state of a DNS calculation. The figure shows that this covariance of the DNS is fairly localised in space; we hypothesise that EENL interactions are important in maintaining the locality of these correlations. This hypothesis is validated by the fact that the correlations delocalise in space as the quasilinear CE2 calculation progresses. Eventually the second cumulant is very delocalised and has a wave-like form. The long-range teleconnections are typical of the quasilinear interaction between waves (in this case thermally driven Rossby waves) and a mean flow.

\subsection{Multiple Jet Solutions}
\begin{figure}
  \centering
  \includegraphics[width=\textwidth]{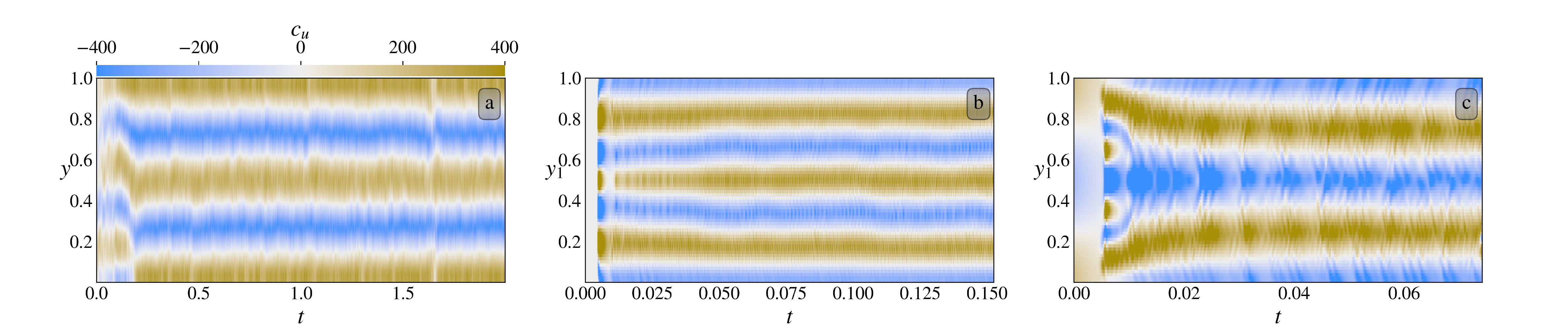}
  \caption{Hovm\"oller diagrams of $\cu$ for run R (a) in DNS showing a 5-jet profile, (b) run R in CE2 showing a stable 7-jet solution when the initial $\cu$ is zero, and (c) run Rb in CE2 with initial $\cu$ biased with a finite-amplitude 3-jet profile. In this case, CE2 latches on to the correct 5-jet solution.}
  \label{fig:hov_run_R}
\end{figure}

\begin{figure}
  \centering
  \includegraphics[width=0.6\textwidth]{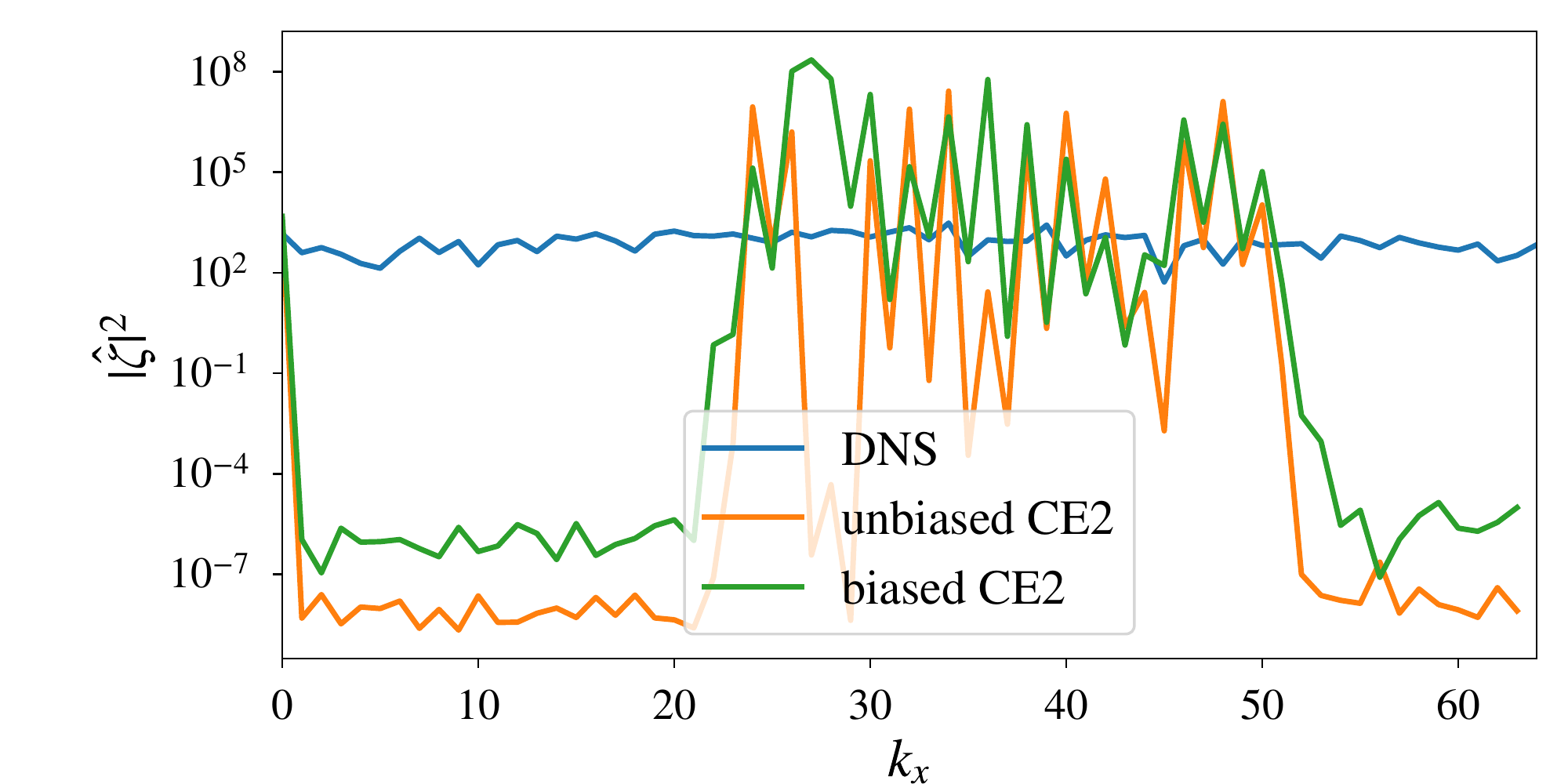}
  \caption{Power spectra of $\zeta$ for run R as a function of $k_x$ for DNS, unbiased CE2, and biased CE2. Note that DNS has energy distributed across many $k$, demonstrating the importance of the EENL. However, both CE2 runs show power only at $k_x 0$ and a band from $20 \lesssim k_x \lesssim 50$ despite the fact that the biased solution gets the correct $\cu$ and the unbiased one does not. Taken together, this suggests that EENL is not crucial for maintaining the mean flow and may only lead to additional dissipation.}
  \label{fig:power_spec_S}
\end{figure}

Parameters may also be selected that yield multiple jet solutions; case R, for which the DNS is shown in Figure~\ref{fig:hov_run_R}(a), is such a case. Here, after an initial transient that takes the form of a six jet solution, a jet merging leads to the solution settling down into a five jet solution. DNS exhibits a solution for the second cumulant that is very localised in space, which is suggested by the broad power spectrum of the DNS shown in Figure~\ref{fig:power_spec_S}. For this case, the maximally ignorant quasilinear CE2 model yields an incorrect seven jet solution. Remarkably, if the CE2 calculation is biased  initially to yield a three jet solution, the solution evolves via a seven jet solution to form a five jet solution of the correct amplitude, obviously driven by the corresponding evolution of the second cumulants; the basin of attraction for CE2 is clearly very sensitive to the initial condition here. The power spectra of the CE2 calculations shown in Figure~\ref{fig:power_spec_S} demonstrate that power is concentrated in a finite number of $k_x$ values indicating a delocalisation of the second cumulant; the second cumulant is again a superposition of wave-like solutions. The precise wavenumbers that receive energy for CE2 are sensitive to the structure of the first cumulant and therefore the initial conditions. Moreover, the  ``zigzag'' structure of the power spectra in CE2 shows that most of the power is trapped within a symmetry subspace and is not efficiently scattered into all modes. Nonetheless we stress that the interactions are sufficient to drive the correct mean flows.  As for Case A, the CE2 solution also tracks the DNS solution well if a ``maximal knowledge'' initial condition is used (not shown).



\subsection{Bursting Dynamics}

\begin{figure}
  \centering
  \includegraphics[width=\textwidth]{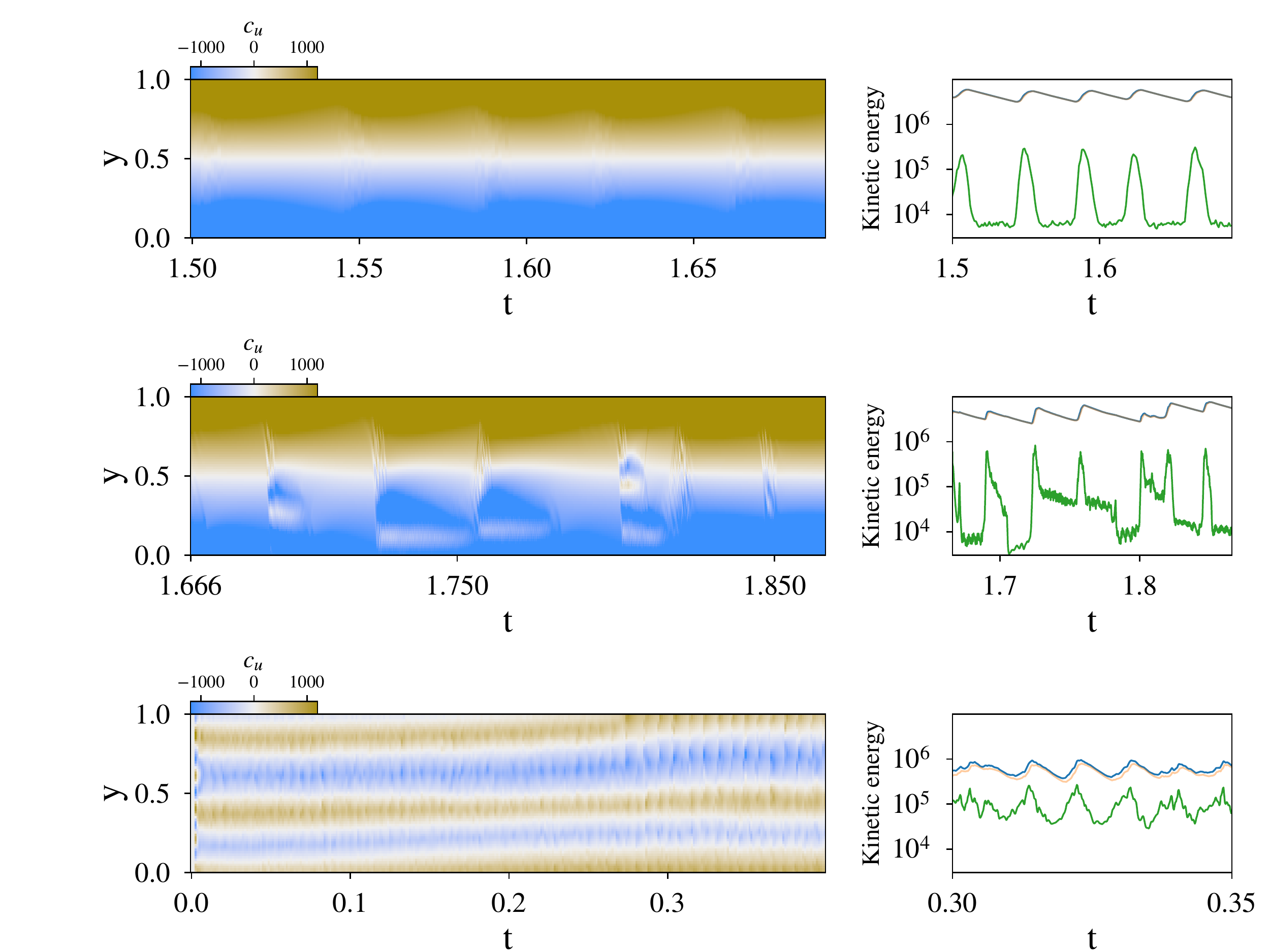}
  \caption{ Left: Run C Hovmuller diagrams of $\cu$. Right: total (blue), zonal (orange), and non-zonal kinetic energies (green). In both columns, top: DNS, middle: ``maximum knowledge'' CE2 initialised from the DNS data in a bursting state, bottom: ``maximum ignorance'' CE2 initialised from a diagonal $\ctt$.}
  \label{fig:run_C}
\end{figure}

The Busse Annulus model is capable of producing extremely complicated nonlinear spatio-temporal dynamics. Perhaps the most nonlinear behaviour exhibited by the DNS model is shown in the top panel Figure~\ref{fig:run_C}(a), which show timeseries of the total, zonal, and non-zonal kinetic energies of the flow and Hovm\"oller plots of the zonally averaged zonal flows for DNS. Here the solution takes the form of a relaxation oscillation; a cycle proceeds as follows. Convection is driven by strong temperature gradients and interacts with the rotation to generate zonal flows (in this case a two-jet solution). As more energy is pumped into the zonal flow, the shear acts so as to switch off the convection (acting as a barrier to transport). This in turn removes the energy source for the zonal flow, which decays on a longer timescale. Once the zonal flow is weak enough convection sets in again and the process is repeated. This type of behaviour has been described in terms of predator-prey dynamics with the convective turbulence taking the role of the prey and the zonal flows acting as a predator. 

Figure~\ref{fig:run_C}(b) shows one of the ``maximum knowledge'' solutions for CE2. Here the solution is started from the DNS solution when it has reached a peak in the zonal flow energy. Remarkably, CE2 is able to continue relaxation oscillations from this state; we believe that this is the first time a quasilinear model has reproduced such complicated behaviour; previously they have been able to generate either the state at the peak or the trough of such a relaxation oscillation --- but not the transition between them \citep{pmt_2019}. If CE2 is started from the trough in the zonal flow energy, similar results are obtained (not shown). The sensitivity to initial conditions for CE2 is, however, demonstrated in Figure~\ref{fig:run_C}(c), which shows the evolution from small amplitude ``maximal ignorance'' initial conditions. This solution clearly does not replicate the correct number of jets --- though it does show some indication of bursting behaviour, as the strength of each zonal jet waxes and wanes in response to the driving.

\subsection{Rank of Second Cumulants}
\label{sec:rank}
One important difference between QL and CE2 simulations is that the former explicitly provides only a single realization of the dynamics, while the latter does not; it provides a description of the low-order statistics of the system and hence can be thought of as averaging over a number of simulations of the system. 
Recently, \citet{nivarti_22}  have shown that an important difference between the two can be quantified as the difference in the ranks of the second cumulants --- that is, the number of non-zero eigenvalues they possess. They have demonstrated that the statistical description CE2 is open to instabilities that can lead to higher rank solutions for the second cumulant than are accessible for QL. Moreover this can lead to different behaviour for CE2 from that found for a single run of a QL system.

\begin{figure}
    \centering
    \includegraphics[width=0.3\textwidth]{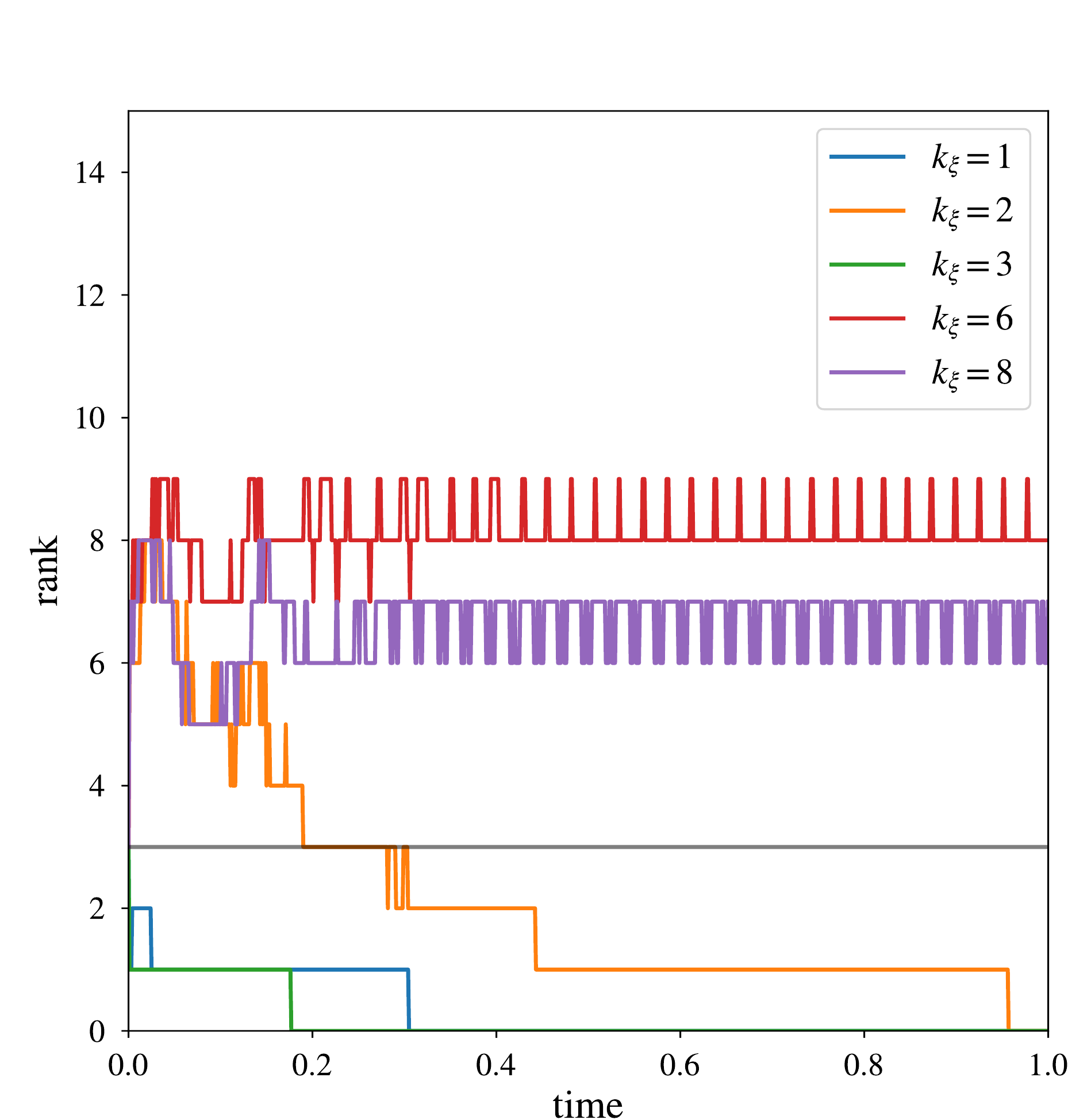}
    \includegraphics[width=0.3\textwidth]{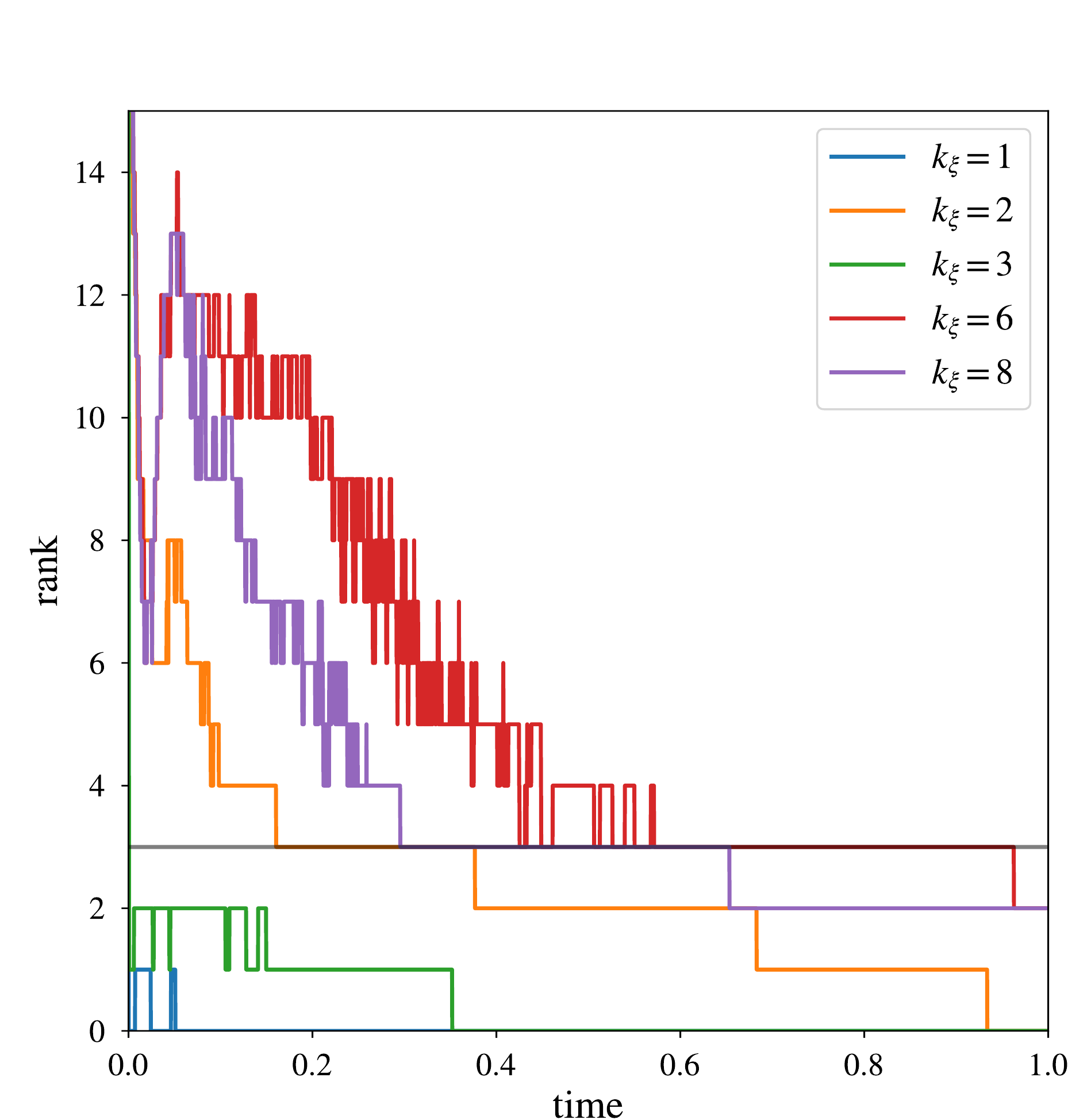}
    \includegraphics[width=0.3\textwidth]{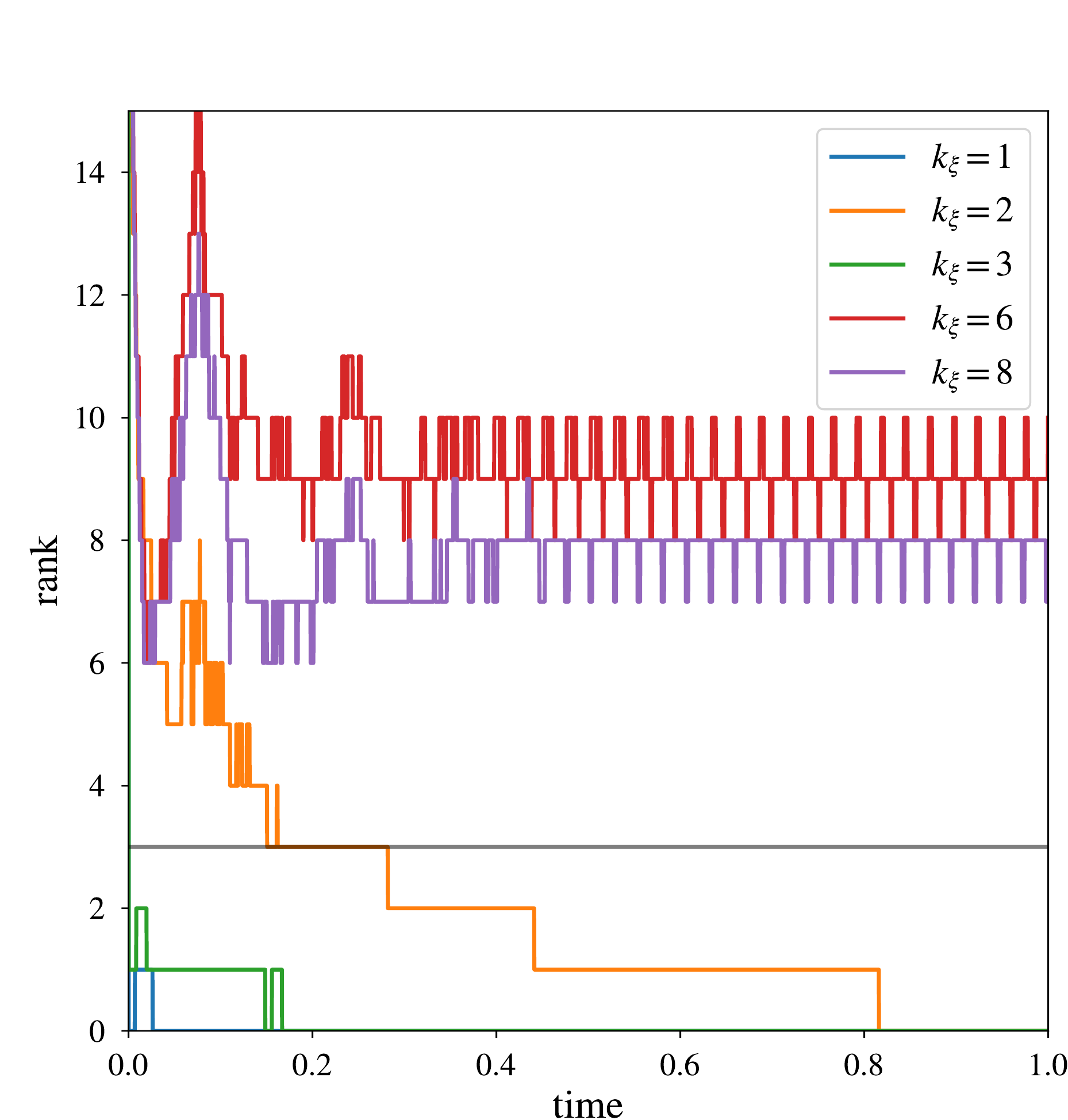}
    \caption{Rank at different zonal wavenumbers as a function of time. (Left) maximum knowledge run A, (middle) maximum ignorance run A, and (right) run A parameters with biased first cumulant.}
    \label{fig:rank_runs}
\end{figure}

Although we do not investigate these instabililities in detail here, it is interesting to note the rank of the modes of the solutions for CE2. Figure~\ref{fig:rank_runs} shows the rank of the modes of the second cumulant as a function of time, for run A and various initial conditions. For a QL system the rank of the solution should be three (note the counting of the rank here is different from that in \citet[][]{nivarti_22}).
For a maximum knowledge initial condition for Run A, when CE2 gives the correct answer, the solution accesses a higher rank solution than QL; note QL obtains the incorrect answer. The solutions clearly undergoes a rank instability to access the correct answer. However, for a maximum ignorance initial condition both QL and CE2 give the incorrect answer. The CE2 solution eventually returns to a low rank solution and reproduces the QL (rather than the correct) dynamics. Finally, for the 
biased initial condition case (with the first cumulant correct but the second cumulant incorrect) the second cumulant remains high rank and one gets the correct behaviour for the first cumulant. We stress again here that in all cases the second cumulant is necessarily delocalised for these quasilinear theories and hence incorrect.

\section{Discussion}
\label{sec:discussion}

In this paper we have examined the effectiveness of the statistical quasilinear theory zCE2 in reproducing the statistics of turbulent solutions of a model of the interaction of rotating convection and zonal flows. This Busse Annulus model exhibits complicated spatio-temporal dynamics including the formation of large-scale zonal jets, multiple zonal jets and even ``predator-prey'' relaxation oscillations between states with strong zonal flows and strong convection.

We show that zCE2 is capable of reproducing even very complicated mean dynamics (such as the driving and decay of the mean flows and modification of the mean temperature profile) though, for this system, it may have multiple stable attractors for any given parameter set. If the system is initiated with \textit{maximal ignorance} then it is possible to fall into the basin of attraction of an attractor that is not preferred by DNS. It is possible to bias the symmetry of the initial condition in order to achieve solutions that mirror the symmetry of the DNS solutions. Furthermore one may use an \textit{maximal knowledge} initial condition where the zCE2 is initiated using the statistics from the saturated state of a DNS run. In this case, zCE2 appears to be extremely successful --- even reproducing the extremely nonlinear relaxation oscillations of the highly driven system.

zCE2 relies on the solution of equations for quantities averaged in the zonal direction, i.e. the mean flows and temperature and the two-point correlation functions.  It is a quasilinear theory in that the evolution equation for the two-point correlation function 
is linear in the two-point correlation function, though there are terms that involve the product of the means and two-point correlation functions. The results presented here indicate that even the predator-prey dynamics is controlled by the interaction of the mean flows and temperature with the fluctuations, with the eddy/eddy $\rightarrow$ eddy nonlinearity (EENL) being of secondary importance. 

It is interesting to speculate further on the role of the missing physics encoded in the EENL. This term in the equations can be modelled by deriving evolution equations for the higher order cumulants, and truncating the hierarchy at third order. This extension to DSS is known as CE3 --- and sometimes CE2.5 when an approximate form of the CE3 equations are utilised \citep{marston_qi_tobias_2019}. However this is a computationally expensive procedure and solution strategies suffer severely from the ``curse of dimensionality''. An interesting question then is what is the best strategy for modelling these higher-order interactions. Of course, if the linear operator provided by the mean flows and other fields is sufficiently non-normal (e.g. if for example the shear is sufficiently strong), then the precise form of this modelling term is not critical (since its role will be to act as a driving term for the non-normal modes of the linear operator). In this case, one may as well model these terms via delta correlated white noise. However it is possible to optimise the form of the noise provided to the linear operator so as to best match the true dynamics. Similar optimisation procedures have been utilised for the form of the (coloured) noise in transition problems \citep[][]{zjg_2017}. One strategy that appeals is to use data-driven methods to ``learn'' the optimal form of the term that should be added to CE2 to best reproduce the dynamics. It is to be hoped (though by no means certain) that this optimal form should be relatively insensitive to the form of the large scales since this dependence should be encoded in the interaction between the first and second cumulants of CE2.   

\backsection[Acknolwedgements]{Computations were performed on the Bates High Performance Computing Center's \emph{Leavitt} system and the NASA High-End Computing Program through the NASA Advanced Supercomputing Division at Ames Research Centre under allocation GID s1647.}
\backsection[Funding]{This work was supported by funding from the European Research Council (ERC) under the EU’s Horizon 2020 research and innovation programme (grant agreement D5S-DLV-786780). 
JSO was supported by NASA LWS grant number NNX16AC92G and Research Corporation Scialog Collaborative Award (TDA) ID\#24231. JBM and SMT are supported in part by a grant from the Simons Foundation (Grant No. 662962, GF).}

\bibliographystyle{jfm}
\bibliography{busse}

\end{document}